\documentstyle[noframe]{mylaa}
\def\cm2{\,{\rm cm^{-2}}}
\def\cm3{\,{\rm cm^{-3}}}

\def\mK{\,{\rm mK}}

\def\pc2{\,{\rm pc^{-2}}}

\def\kms{\,{\rm {km\,s^{-1}}}}
\def\Msun{\,{\rm M_\sun}}

\def\C34S{\,{\rm C^{34}S}}
\def\13CS{\,{\rm ^{13}CS}}

\def\HII{\mbox{H\,{\sc ii}}}
\def\HI{\mbox{H\,{\sc i}}}
\def\Tmb{$T_{\rm mb}$\,}

\def\Emb{$\eta_{\rm mb}$}
\def\Efss{$\eta_{\rm fss}$}
\def\to {$\rightarrow$}
\def\sun{\odot}

\def\etal{{et al.} }

\begin{document}
 
\thesaurus{}
 
\title{CO~J\,=\,2\to1 observations of the nearby galaxies Dwingeloo~1
and Dwingeloo~2}
 
 
\author{R.P.J. Tilanus \inst{1,2}  \and W.B. Burton \inst{3}} 
\institute{Netherlands Foundation for Radio Astronomy, P.O. Box 
2, 7990 AA Dwingeloo, the Netherlands 
\and Joint Astronomy Centre, 660 N. A'ohoku Place, Hilo, Hawaii, 
96720, USA 
\and Sterrewacht Leiden, P.O. Box 9513, 2300 RA Leiden, the 
Netherlands} 

\offprints{R.P.J.\ Tilanus, Joint Astronomy Centre, Hilo} 
  
\date{Received date; Accepted date} 
 
\maketitle 

\begin{abstract}  We present an exploratory survey of the central 
regions of the nearby spiral galaxies Dwingeloo~1 and Dwingeloo~2 made
in the CO~J\,=\,2\to1 line at 230 GHz using the James Clerk Maxwell
Telescope. In Dwingeloo~1, which is probably the nearest grand--design
barred spiral, the CO emission was sampled along a cross with its
major axis aligned with the bar. The CO emission is an order of
magnitude weaker (peak \Tmb $\approx$ 70\mK) than in its neighbours
IC~342 and Maffei~2 and appears to be confined to the nucleus, bar,
and spiral arms. In Dwingeloo~2, a small system quite likely to be a
companion of Dw\,1, a 3-sigma upper limit of 40$\mK$ for the
CO~J\,=\,2\to1 emission from the nucleus was obtained.

 
\end{abstract} 

\section{Introduction} 

Dwingeloo~1 is a nearby barred spiral system discovered in the \HI\
line by Kraan-Korteweg \etal (1994) during the Dwingeloo Obscured
Galaxies Survey for galaxies hidden in the Zone of Avoidance.
Subsequent optical and infrared observations by Loan \etal (1996),
Aspin \& Tilanus (priv. comm.), and McCall \& Buta (1996), as well as
interferometric \HI\ observations by Burton \etal (1996), describe
Dw\,1 as of morphological type SBb at an inclination of
$i=50^{\circ}$, with a systemic velocity with respect to the LSR of
$110\pm0.4\kms$, and at a distance estimated to be 3 Mpc.  The small
galaxy Dwingeloo~2 was discovered by Burton \etal (1996) in the
primary beam of the Westerbork observations as pointed towards Dw\,1.
In view of its angular and kinematic proximity to Dw\,1, and in view
of distortions in its velocity field, it seems likely that Dw\,2 is a
companion of the larger system.  The two galaxies are probably members
of the group containing Maffei~1 \& 2 and IC$\,$342 and may influence
the motions and morphology of that group and, by the group
collectively, of our own Local Group.

There is yet but little information on the star-forming potential of
Dw\,1 and Dw\,2.  Loan \etal reported numerous \HII\ regions in Dw\,1.
Li \etal (1996) detected what they characterised as weak
CO~J~=~1\to0 emission from the core of Dw\,1.  The ratio of
CO--to--\HI\ emissivity varies widely in galaxies, especially in
barred galaxies (see e.g. Young \& Scoville 1991).  But Li \etal were
not able to conclude from their single spectrum if the ratio in Dw\,1
is unusally low, or if the galaxy belongs amongst those that show a
central hole in the molecular gas distribution.  The results of the
exploratory survey reported here indicate that Dw\,1 is not a strong
CO emitter, but that both its CO emission and the CO--to--\HI\ ratio
fall within the range exhibited by ordinary galaxies. Rather than
showing a central hole the molecular emission from the core is in fact
more intense than that from beyond the core. We were unable to detect
any CO emission from Dw\,2.

\section{Observations} 

The observations of Dw\,1 and Dw\,2 were made with the James Clerk
Maxwell Telescope atop Mauna Kea in July and August of 1995. We used
the A2 receiver which employs a lead--alloy SIS mixer and a
Carlstrom--Gunn local oscillator; it has a noise temperature of about
95\,K.  The typical single--sideband system temperature is about
350\,K at 230\,GHz.  The DAS digital autocorrelator backend was used
with 2048 channels and configured for a total bandwidth of 500\,MHz;
during data reduction the velocity resolution was subsequently
smoothed to 5\,MHz, about 6.5$\kms$.

The size of the beam of the JCMT 15--m attenna at the frequency
(230.5380 GHz) of the CO~J~=~2\to1 transition is 21$\arcsec$.  The
absolute pointing of the observations was good to about 2.5$\arcsec$
rms; relative pointing was better than this. The Dw\,1 spectra were
obtained by position--switching the telescope to reference positions
which were in general also located on that galaxy, but at the opposite
side of the bar.  Since galactic rotation shifts the signals from the
two sides well apart in frequency, this strategy seemed a valid one to
optimize the use of telescope time. In order to guard against
accidental subtraction of emission due to an overlapping signal at the
reference position, several different reference positions were
commonly used with each `on' position, and the profiles from each set
carefully compared. The spectra were calibrated in units of the
main--beam brightness temperature (\Tmb)\footnote{All reported CO
intensities and derived molecular gas parameters are based upon \Tmb},
and corrected for sideband gains, for atmospheric emission in both
sidebands, and for telescope efficiency. The rms surface accuracy of
the JCMT is of order 30$\,\mu$m.  At the CO~J\,=\,2\to1 frequency,
the forward-scattering and spillover efficiency, \Efss, has a value of
0.8; the main-beam efficiency, \Emb, is 0.69.  Residual baseline
offsets were corrected by polynomal baseline removal.

We observed a total of 15 positions in Dw\,1 with the inner 11 on a
cross which had 34$\arcsec$ spacings and was rotated over a position
angle of 118$^{\circ}$ to align it with the central bar of Dw\,1.  The
grid corresponds to an ($\alpha,\delta$) cell of $30\arcsec$ by
$16\arcsec$.  The center of Dw\,1, our (0,0) position, was taken as
$\alpha$(1950) = 02$^{\rm h}$53$^{\rm m}1\fs0$, $\delta$(1950) =
58$\degr$42$\arcmin$38$\arcsec$ $(l,b = 138\fdg52, -0\fdg11)$ at an
LSR velocity of 110$\kms$, corresponding to the values determined
initially by Kraan--Korteweg \etal (1994) and confirmed by the WSRT
observations by Burton \etal (1996). The remaining 4 positions were
located at larger distances at $\pm150\arcsec$ and $\pm240\arcsec$
along the direction of the bar.

\begin{figure*} 
\picplace{13.0cm} 
\includegraphics{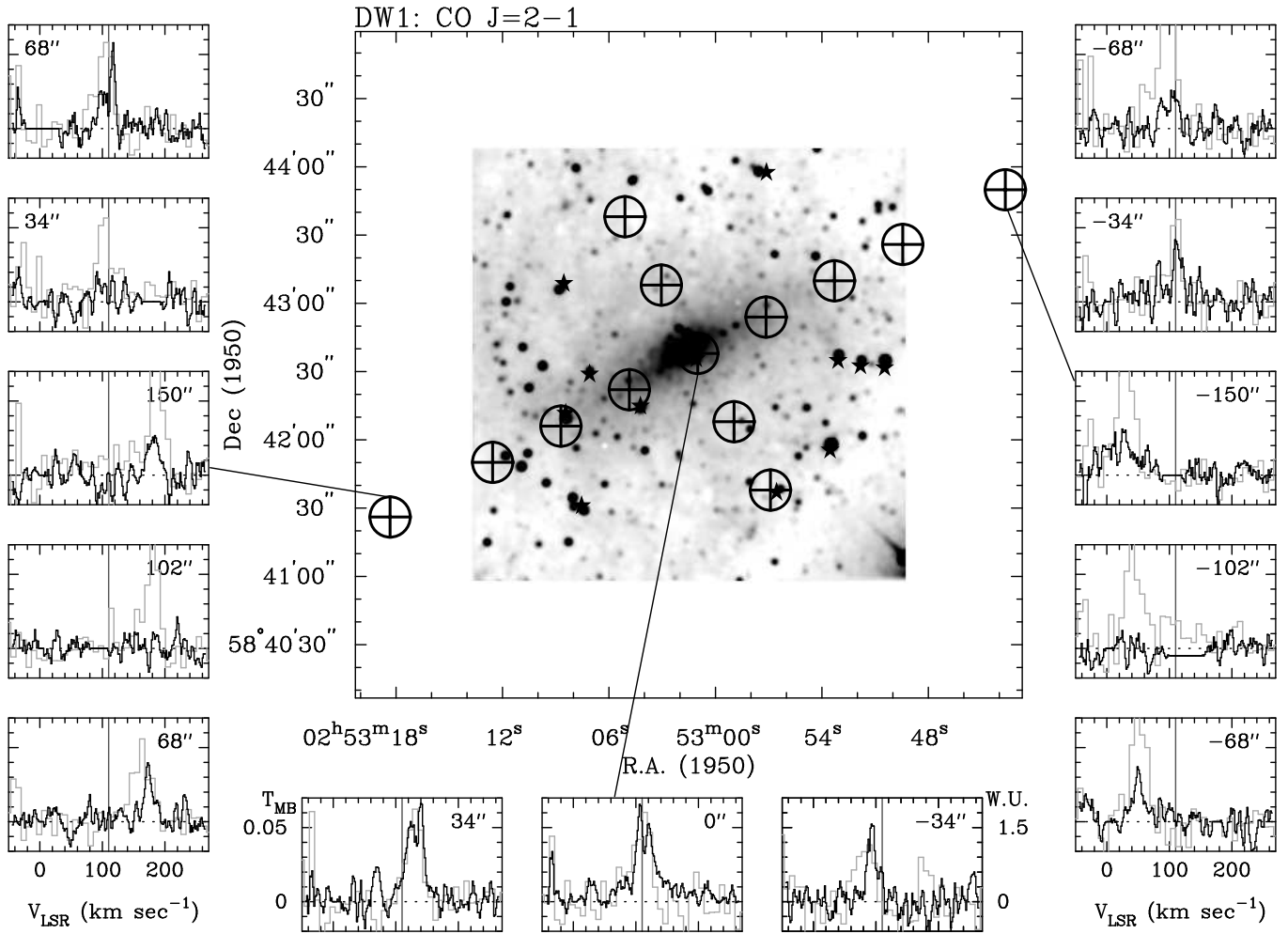}
\caption[]{CO J = 2\to1 emission spectra as observed toward 
Dwingeloo~1 (dark lines) shown superposed on the \HI\ spectra 
observed by Burton \etal (1996).  The central image of the 
galaxy was observed by Aspen and Tilanus in the K band, using the 
UKIRT facility.  The circles on the K--band image indicate the 
positions observed in the CO line; the size of the circles 
corresponds to the size of the JCMT beam at 230\,GHz. Note that 
the upper two spectra plotted on either side of the K--band image 
were observed perpendicular to the bar.  The spectral intensities are 
given in units of \Tmb. The spectra have varying noise levels due 
to varying weather conditions and integration times. The vertical 
line indicates the systemic LSR velocity of Dw\,1, 110$\kms$.  
The appropriate distance in arcseconds from the center of Dw\,1 
is indicated in each panel. \\} \end{figure*}

     The JCMT CO spectra are shown in Figure~1, superposed on the \HI\
spectra observed by Burton \etal (1996) using the WSRT.  The \HI\ data
were convolved to the spatial resolution of the current CO data prior to 
the extraction of the spectra.  Shown in the center of the figure is a
UKIRT K--band image of Dw\,1 observed by Colin Aspin and Remo Tilanus
and provided to us in advance of publication.  Of the 15 positions
searched for CO, 5 are located on the central bar.  CO emission was
detected at most of the positions observed with peak main--beam
brightness temperatures of \Tmb $\approx$ 70\mK.  The kinematics
revealed by the CO data are consistent with those seen in \HI\ at
the positions where both tracers were detected.  It is interesting to
note that the strength of the CO line relative to that of the \HI\
line decreases with increasing distance from the nucleus; thus Dw\,1
is generally not a strong molecular emitter, nor is it a galaxy with an
anomalously low molecular component in the core.

    The sky conditions on successive nights at the JCMT were 
quite variable, and thus so were the integration times invested 
in each observation; the rms noise level varies from profile to 
profile.  The spectra have been smoothed in velocity to a 
resolution of 6.5 $\kms$.  The spectra have also been 
interpolated across a velocity range extending from $-5\kms$ to 
$+10\kms$ in order to blank out some CO emission contributed by a 
foreground molecular cloud located in the Milky Way; this 
foreground emission was also evident near zero velocity in the 
spectrum observed by Li \etal (1996).  We cannot rule out 
additional contamination from low level Milky Way emission 
at velocities less than $-5\kms$.  In addition, the spectral 
values have been blanked at those velocities where a strong 
emission feature was present at the reference position. 

     It is necessary to have accurate positional information to 
assist superposition of data garnered at different wavelengths.  
Table 1 lists astrometric positions of stars located around the 
direction of Dw\,1, as determined from POSS prints. The accuracy 
of the positions is about one second of arc. 

\begin{table} 
\caption[]{Positions of stars around Dw\,1} 
\begin{flushleft} \begin{tabular}{rrr} \hline 
 id & $\alpha(1950)$ & $\delta(1950)$ \\ 
& ($^{\rm h}\, ^{\rm m}\, ^{\rm s}$) & ($^{\degr}\, ^{\arcmin}\, 
^{\arcsec}$)  \\ 
\hline \\ 
  1 & 02 53 08.54 & +58 43 08.6 \\
  2 & 02 53 07.09 & +58 42 28.8 \\
  3 & 02 53 08.44 & +58 42 11.6 \\
  4 & 02 53 04.23 & +58 42 14.8 \\
  5 & 02 53 01.08 & +58 42 35.6 \\
  6 & 02 52 53.09 & +58 42 34.8 \\
  7 & 02 52 51.82 & +58 42 32.5 \\
  8 & 02 52 50.46 & +58 42 31.6 \\
  9 & 02 52 53.54 & +58 41 55.4 \\
 10 & 02 52 56.55 & +58 41 36.9 \\
 11 & 02 53 07.55 & +58 41 31.0 \\
 12 & 02 52 57.13 & +58 43 57.5 \\
 13 & 02 52 21.07 & +58 43 44.2 \\
\hline 
\end{tabular} 
\end{flushleft} 
\end{table} 

\begin{table*} \caption[]{Column densitites in Dw\,1 as observed in 
the CO~J\,=\,2\to1 transition and as estimated for H$_{2}$} 
\begin{flushleft} \begin{tabular}{rrrrrrl} \hline & $\Delta\alpha$ 
& $\Delta\delta$ & distance & $\int T_{\rm mb}$d$V$ & 
$N$(H$_{2}$)  & remarks \\ 
& ($\arcsec$) & ($\arcsec$) & ($\arcsec$) & (K $\kms$)  & 10$^{20}$ 
cm$^{-2}$ &  \\ 
\hline 
\multicolumn{2}{l}{Along bar:} & & & & & \\ 
 &    0  &   0  &    0 &  2.69  &  10.0  & \\
 &   30  & -16  &   34 &  2.10  &   7.9  & \\
 &  -30  &  16  &  -34 &  1.00  &   3.8  & \\
 &   60  & -32  &   68 &  0.70  &   2.6  & \\
 &  -60  &  32  &  -68 &  0.73  &   2.7  & \\
 &   90  & -48  &  102 &  0.07  &   0.3  & \\
 &  -90  &  48  & -102 &  0.05  &   0.2  & \\
 &  132  & -70  &  150 &  0.54  &   2.0  & \\
 & -132  &  70  & -150 &  0.65  &   2.4  & contaminated by MW? \\ 
 &  212  & -113 &  240 &  0.05  &   0.2  & \\
 & -212  & 113  & -240 &  0.41  &   1.5  & contaminated by MW? \\ 
\multicolumn{2}{l}{Perpendicular bar:} & & & & & \\ 
 &   16  &  30  &   34 &  0.40  &   1.5  & \\
 &  -16  & -30  &  -34 &  1.10  &   4.1  & \\
 &   32  &  60  &   68 &  0.88  &   3.3  & \\
 &  -32  & -60  &  -68 &  0.63  &   2.4  & \\
\hline 
\end{tabular} 
\end{flushleft} 
\end{table*} 

\section{Discussion}

The overall appearance of the CO J = 2\to1 spectra supports the
conclusion of Li \etal (1996) that the CO emission from the nucleus of
Dw\,1 is relatively weak, but only in the sense that it is much weaker
(by an order of magnitude) than the strong CO emission from its
neighbours, IC~342 and Maffei~2. Li \etal speculated that the CO
emission from the nucleus of Dw\,1 might not be typical of the galaxy
as a whole, but that Dw\,1 might be characterised by a central hole in
the distribution of molecular gas.  The present observations show,
however, that the CO emission from the nuclear region is actually
substantially more intense than that detected along the bar or from
the spiral arms.  Along the bar, there is a general decline of the CO
emission towards the start of arms at the end of the bar.  Just beyond
the bar--arm intersections in the interarm regions the CO emission
becomes undetectable. The profiles observed perpendicular to the bar at
68$\arcsec$ distance from the nucleus, especially north, show some
indication that CO emission is associated with the arms. Further
support for this comes from the profiles observed along the bar
at ${\pm}150\arcsec$ which may be located along the continuation of
the spiral arms visible in the K--band image and show emission at the
expected velocities. In contrast, the two positions closer in
(${\pm}102\arcsec$), which are clearly interarm locations, show no
detectable CO emission.  However, the coverage of the current
observations is insufficient to support a firm statement on the
detailed association of CO emission with the spiral arms.

These conclusions are supported by Figure~2, which shows the
integrated CO emission as a function of galactocentric radius.
Figure~2 also shows the radial distribution of \HI\ from the
observations of Burton \etal (1996).  The CO--to--\HI\ ratio by mass
observed in many other galaxies covers a wide range of values.  The
lower panel of Figure~2 shows that in Dw\,1 the \HI\ and H$_2$ masses
are rather similar outside of the nucleus, where H$_2$ dominates \HI\
by a factor of about 2.5. The relative amount of atomic and molecular
gas in Dw\,1 is thus not very different from the situation pertaining
in the more familiar galaxy M\,33.

\begin{figure} 
\picplace{8.5cm}
\includegraphics{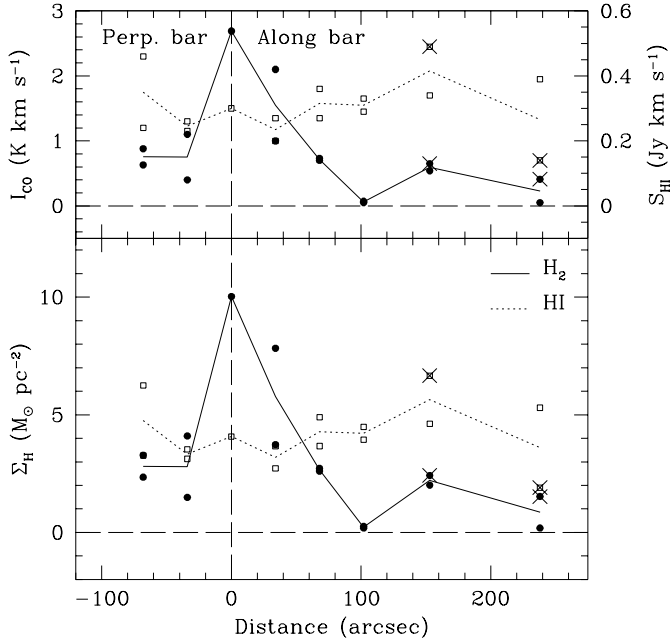}
\caption[]{Distribution of the integrated CO~J~=~2\to1 and \HI\ 
emission along, and perpendicular to, the bar in Dwingeloo~1. In 
the lower panel, the observed emissivities have been converted to 
deprojected surface density units corrected for an inclination of 50$\degr$.
Positions marked with a cross suffer from contamination by CO emission
from our own Galaxy, which, although subtracted as well as possible,
does render a reliable baseline determination difficult. The angular
distances indicated are not corrected for projection due to the inclination 
of Dw\,1. The molecular and atomic components are approximately equal in
surface density away from the nucleus, but the molecular gas dominates
in the core. \\}
\end{figure} 

   Figure~3 shows the CO spectrum observed toward the nucleus of 
Dwingeloo~1, together with a three--component Gaussian 
decomposition of the observation. The integrated intensity of 
$I_{\rm CO} = 2.7$ K[\Tmb]$\kms$ is about 45\% higher than
the value for the integrated CO~J\,=\,1\to0 line found by Li \etal 
(1996) with a 55$\arcsec$ beam (and after multiplying our number 
by \Emb/\Efss$ = 0.86 $ for the comparison). Their profile, which 
was smoothed to a resolution of 15$\kms$, does not show the 
double--peaked structure which our observation does. The two 
principal Gaussians peak at 106.5$\kms$ and 122.0$\kms$, 
respectively, suggesting a systemic LSR velocity of about 
114$\kms$, which is a velocity close to the fitted center of the 
broad component, at 118$\kms$, and which agrees well with the 
center of a single--component Gaussian fit, at 115$\kms$. The 
single--component FWHM width is about 40$\kms$. The +34$\arcsec$
profile also is double--peaked like the nuclear profile.
Most likely these spectra show non-circular streaming motions 
of the gas caused by the presence of a bar. The \HI\ observations 
by Burton \etal (1996) place the dynamic center of the galaxy
5 arcseconds east and south of our (0,0) position.

\begin{figure} 
\picplace{4.2cm} 
\includegraphics{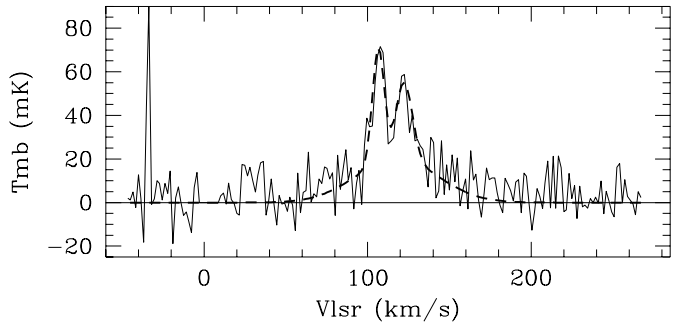}
\caption[]{Spectrum of the CO~J\,=\,2\to1 emission observed toward 
the nucleus of Dwingeloo~1, shown superposed on a three--component
Gaussian decompostion. The amplitudes, the full widths at half
maximum, and the central LSR velocities of the three components are,
respectively, (53, 8.4, 106.5), (36, 11.9, 122.0), and (20, 58.7,
118.0) [mK,$\kms$,$\kms$]. \\} 
\end{figure} 

If we adopt a value of the CO(2\to1)/CO(1\to0) intensity ratio of 0.8
as found for the Milky Way by Sanders \etal (1992) and the
CO(1\to0)--to--H$_{2}$ conversion factor of $3 \times 10^{20}$
H$_{2}\,$cm$^2$ $({\rm K} \kms)^{-1}$ advocated for the Milky Way by Scoville
\etal (1987) and by Bloemen \etal (1986), we find that the average
H$_{2}$ column density\footnote{For details on the derivation of the
numbers presented see the appendix to Kenney \& Young (1989) with
$\chi = 3 \times 10^{20}$ and the assumed CO(2\to1)/CO(1\to0)
intensity ratio of 0.8} within the central 21$\arcsec$ of Dw\,1 is
$1.0 \times 10^{21}$ N(H$_{2}$)\,cm$^{-2}$ corresponding to a surface
density of $16 \Msun\pc2$ and a total H$_{2}$ mass of $1.2 \times
10^{6} ({\rm D \over 3\,Mpc}) \Msun$. Correcting for an inclination of
50$\degr$, one finds a deprojected average H$_{2}$ surface density
and total H$_{2}$ mass for the central 300\,pc of Dwingeloo~1 of $10
\Msun\pc2$ and $7.3 \times 10^{5} ({\rm D \over 3\,Mpc}) \Msun$,
respectively.

We also attempted to detect CO~J\,=\,2\to1 emission from Dwingeloo~2.
The small total velocity width of Dw\,2 shown in the \HI\ data of
Burton \etal (1996), its small size (assuming it is located at the
same distance as Dw\,1), and the fact that it has been detected only
at a weak level in the infrared, together lead one to expect that the
CO signal, if detected at all, would be very weak.  We confined the
JCMT observations to integrations towards the nucleus of Dw\,2, but
were not able to detect CO emission above a 3--$\sigma$ threshold of
40\,mK.

\section{Conclusions} 

The molecular emission observed from the nearby galaxy Dw\,1 in the
CO~J\,=\,2\to1 line is relatively weak compared to its neighbours
IC\,342 and Maffei~2, but is in the normal range for an ordinary
galaxy.  The emission could be followed in the JCMT observations
reported here over the extent of the bar and into the spiral arms
emanating from the ends of the bar. Thus it seems that Dw\,1 does not
belong to the category of barred systems with a central hole in the
molecular distribution; the molecular emission is comparable
throughout the galaxy, although relatively more intense in the core
than in the bar or spiral-arm regions.  Outside of the nucleus, where
H$_{2}$ dominates \HI\ by a factor of about 2.5, the atomic and
molecular masses are rather similar.  We were unable to detect CO
emission from the nucleus of Dw\,2; this negative result is not
unexpected for a dwarf companion system.

\acknowledgements

The James Clerk Maxwell Telescope is operated by the Observatories on
behalf of the Particle Physics and Astrophysics Council of the United
Kingdom, the Netherlands Foundation for Research in Astronomy, and the
National Research Council of Canada.  We are grateful to Chris Dudley
(IfA, University of Hawaii) for help with the stellar astrometry, to
Colin Aspin (UKIRT) for agreeing that we use the Aspin/Tilanus K--band
image of Dwingeloo~1, and to Marc Verheijen (Groningen) for providing
the WSRT \HI\ spectra in the form used here in Figure~1. We also thank
an anonymous referee for constructive comments.)

\end{document}